\documentclass{article}
\usepackage{spconf,amsmath,graphicx}
\usepackage{multirow}
\usepackage{tabularx}
\usepackage{dsfont}
\usepackage{overpic}
\usepackage{enumitem} %< control spacing in itemize/enumerate/...
\usepackage{overpic} %< add raw math symbols to figures
\usepackage{color}
\usepackage{pifont}
\usepackage{amsfonts}
\usepackage{booktabs}
\usepackage{tabularx}

\definecolor{turquoise}{cmyk}{0.65,0,0.1,0.3}
\definecolor{purple}{rgb}{0.65,0,0.65}
\definecolor{dark_green}{rgb}{0, 0.5, 0}
\definecolor{orange}{rgb}{0.8, 0.6, 0.2}
\definecolor{red}{rgb}{0.8, 0.2, 0.2}
\definecolor{darkred}{rgb}{0.6, 0.1, 0.05}
\definecolor{blueish}{rgb}{0.0, 0.3, .6}
\definecolor{light_gray}{rgb}{0.7, 0.7, .7}
\definecolor{pink}{rgb}{1, 0, 1}
\definecolor{greyblue}{rgb}{0.25, 0.25, 1}

\newcommand{\PreserveBackslash}[1]{\let\temp=\\#1\let\\=\temp}
\newcolumntype{C}[1]{>{\PreserveBackslash\centering}p{#1}}
\newcolumntype{R}[1]{>{\PreserveBackslash\raggedleft}p{#1}}
\newcolumntype{L}[1]{>{\PreserveBackslash\raggedright}p{#1}}

\newcommand{\Fig}[1]{Fig.~\ref{fig:#1}}

\newcommand{\Tab}[1]{Tab.~\ref{tab:#1}}

\usepackage{blindtext}

\renewcommand{\paragraph}[1]{\vspace{1em}\noindent\textbf{#1}.}

\newcommand{\eg}{\textit{e.g.~}}

\usepackage{subcaption}
\usepackage{amsmath}
\usepackage{cite}

\def\shortmethod{MoLE}
\def\fullmethod{Mixture-of-Language-Experts}
 
\title{M\lowercase{O}LE : Mixture of Language Experts \\for Multi-Lingual Automatic Speech Recognition}
\name{Yoohwan Kwon, Soo-Whan Chung}
\address{NAVER Cloud, South Korea}
\begin{document}
%\ninept
%
\maketitle
\begin{abstract}
Multi-lingual speech recognition aims to distinguish linguistic expressions in different languages and integrate acoustic processing simultaneously.
In contrast, current multi-lingual speech recognition research follows a language-aware paradigm, mainly targeted to improve recognition performance rather than discriminate language characteristics.
In this paper, we present a multi-lingual speech recognition network named \fullmethod~(\shortmethod), which digests speech in a variety of languages.
Specifically, \shortmethod~analyzes linguistic expression from input speech in arbitrary languages, activating a language-specific expert with a lightweight language gating network.
The gating network not only activates experts, but also estimates the reliability of the activation.
Based on the reliability, the activated expert and the language-agnostic expert are aggregated to represent language-conditioned embedding for efficient speech recognition.
Our proposed model is evaluated in 5 languages scenario, and the experimental results show that our structure is advantageous on multi-lingual recognition, especially for speech in low-resource language.

% The tokenizer not only activates experts, but also provides the reliability degree of the activation.
%\jee{The tokenizer not only activates experts, but also estimates how reliable the activation is.}
% \jee{
% Successful multi-lingual speech recognition systems demand concurrent modelling of both linguistic and acoustic information. 
% However, the majority of recent research literature leverages external language identification models; they rather focus on lowering the metric without explicit modelling of language discriminative techniques. % with
% }
\end{abstract}

\section{Introduction}
\label{sec:intro}

\begin{figure*}[t]
\centering
     \begin{subfigure}[b]{0.48\linewidth}
         \centering
         \includegraphics[width=\textwidth]{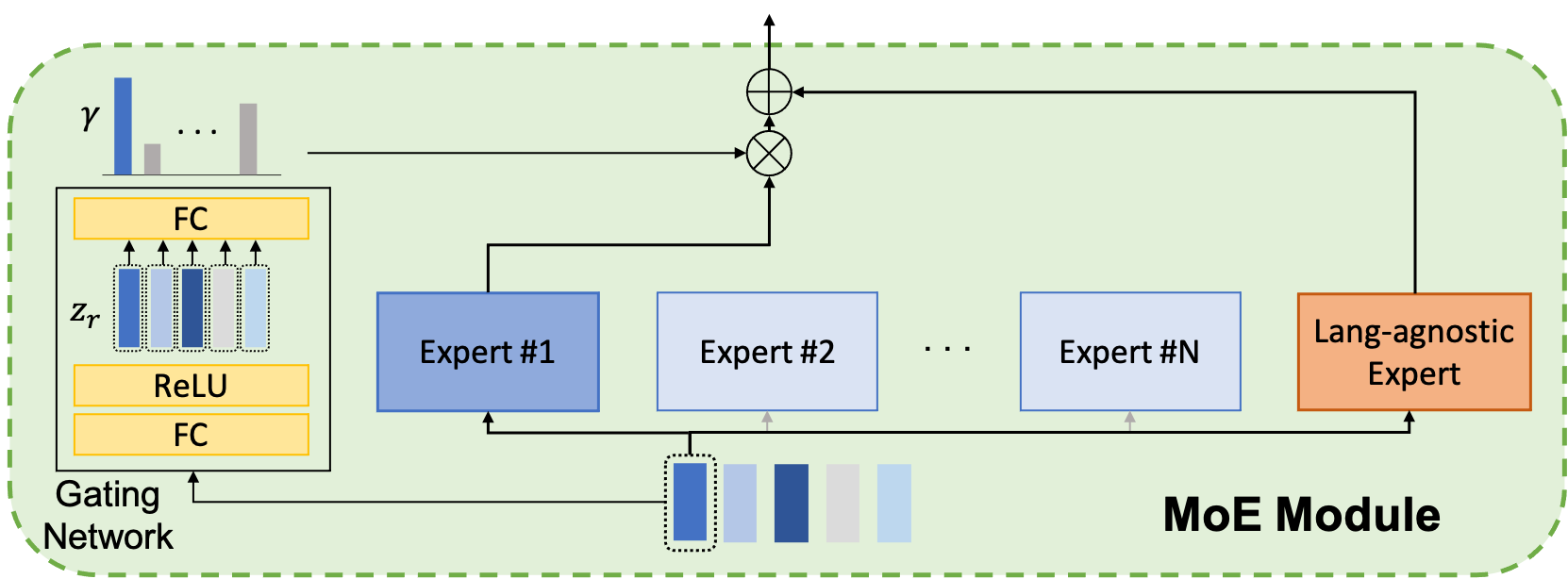}
         \caption{MoE Module}
\vspace{-5pt}
         \label{fig:MoE_Module}
     \end{subfigure} 
     \hfill
     \begin{subfigure}[b]{0.48\linewidth}
         \centering
         \includegraphics[width=\textwidth]{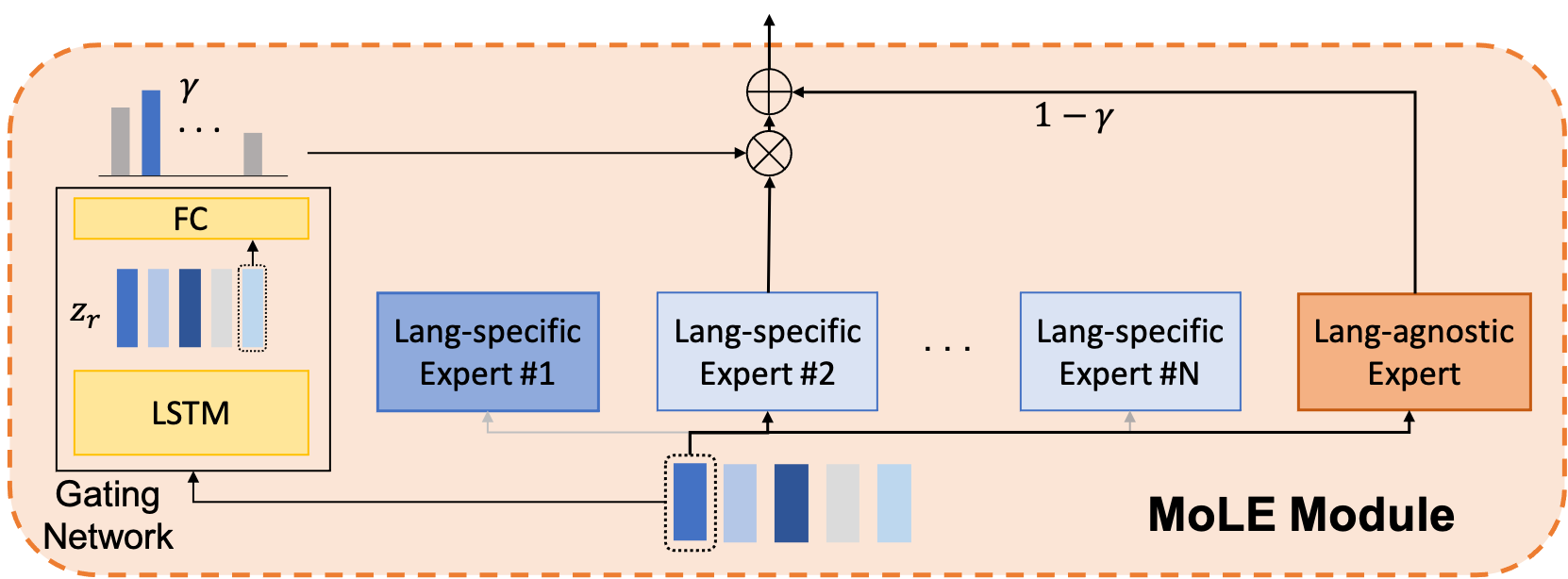}
         \caption{MoLE Module}
\vspace{-5pt}
         \label{fig:MoLE_Module}
     \end{subfigure}    
     \caption{Illustration of the expert blocks. (a) Mixture-of-Experts (MoE), (b) Mixture-of-Language-Experts (MoLE)}
     \label{fig:modules}
    \vspace{-6pt}
\end{figure*}
Humans' speech and conversation have been written in text, and they have become the basis of the plethora of data these days.
Recently, many works have explored linguistic expression in speech signals and converted them into text sequences automatically.
% Although it had been a difficult task to solve due to acoustic variations, deep learning has accelerated the advances of speech-to-text techniques, and a large amount of speech datasets on the Internet has led to significant improvement of its performance.
Although it had been a difficult task to solve due to acoustic variations, deep learning has accelerated the advances of speech-to-text techniques.
Also large-scale speech datasets on the Internet enabled significant performance improvement

In the past, ASR has required phonetic timestamps aligned to speech signals as well as text annotations, and it has consumed a large amount of costs on the annotations.
The connectionist temporal classification~(CTC)~\cite{graves2006connectionist} learning method has overcome this problem using an effective forward-backward algorithm.
Besides, the advent of learning strategies, there has been large progress in the model view, including sequence-to-sequence models~\cite{chorowski2015attention,chan2016listen,dong2018speech} and RNN-transducer models~\cite{graves2012sequence,zhang2020transformer,gulati2020conformer}.
% and transformer-based speech analysis models~\cite{dong2018speech,gulati2020conformer}.

Although there have been dramatic improvements in ASR, there still remain several challenges.
The problem that we address in this paper is the multi-lingual ASR, which distinguishes language-related expression and integrates acoustic processing, simultaneously.
Multi-lingual ASR has two major difficulties compared to mono-lingual counterpart.
First, if the ASR model cannot detect language well, the model resorts to wrong vocabulary sets, which causes catastrophic results.
Many works have avoided this problem by providing explicit language labels, but this solution requires an external language identifier.
Second, unbalanced speech data between different languages biases the training to more dominant language with sufficient data while the others are left in an under-fitting state.
Several previous works handled these difficulties with various approaches; \eg data efficiency~\cite{alumae2016improved,sercu2016very}, multi-task learning~\cite{hou2020large,7084614}, adapter network~\cite{kannan2019large,winata2020adapt,hou2021meta}, and meta-learning~\cite{hsu2020meta,xiao2021adversarial}.

In this paper, we propose a novel multi-lingual ASR model, named \textit{\fullmethod~(\shortmethod)}, which perceives speech in a variety of languages with implicit language router and converts it to text sequences.
In \shortmethod, we leverage a variation of a mixture-of-experts~(MoE) network~\cite{you2021speechmoe,zhou2022configurable,gaur2021mixture,lepikhin2020gshard} and a switch transformer~\cite{Fedus2021SwitchTS,kumatani2021building}, which have shown great performances in various fields.
Our network consists of $N+1$ experts within a {\shortmethod} block, including $N$ \textit{Language-specific Experts~(LsEs)} and a \textit{Language-agnostic Expert~(LaE)}.
We also add a LaE to make up for the confusion of the language gating network and to support generic acoustic modeling.
% Our proposed method is efficient and suitable for the multi-lingual ASR task because it doesn't require an external language classifier since it implicitly distinguishes language using a gating network.
Our proposed model is efficient and suitable for the multi-lingual ASR task since it implicitly distinguishes language types without the help of an external language classifier.
Moreover, even if the number of languages increases, the increase in computational complexity in the overall model is extremely minimal because the gating network activates only one LsE per input signal.
In experiments, we confirm the superiority of the proposed model compared to both the vanilla transformer model and the MoE-based transformer model.
The effectiveness of our model is more pronounced in the results of low-resource language speech recognition.

\section{Mixture-of-Experts}
\label{sec:related}

Deep MoE~\cite{eigen2013learning} has been proposed to increase the representation power of a shallow network architecture, by placing multiple sub-modules or experts in a layer.
The experts are dynamically weighted by a gating network, thus it provides sparsity on the network and re-calibrates the network output.
The output of a MoE layer is described as,
\vspace{-3pt}
\begin{equation}
    y=\sum_{i=1}^N p_i(x) e_i(x),
\vspace{-3pt}
\end{equation}
where $N$ is the number of experts, $x$ is the input, $p_i(\cdot)$ and $e_i(\cdot)$ indicate the $i$-th posterior of the gating network and the expert network, respectively. 
The conventional MoE approaches benefit model scaling and improve performances at the cost of more computations.
% Afterwards, sparse MoE has overcome the problem by selecting a few experts among them, and it has enabled placing a large complexity layer as an expert.

Recent works have incorporated the MoE structure with a transformer network~\cite{lepikhin2020gshard, Fedus2021SwitchTS}.
In~\cite{Fedus2021SwitchTS}, they proposed a sparse MoE-based transformer.
The sparse MoE has solved the increasing computation problem by leveraging only limited number of experts at a time regardless of the total number of experts.
The modified gating network selects experts with the highest scores rather than weighting experts which can be described as,
\vspace{-3pt}
\begin{equation}
    y=\sum_{i=1}^N p_i(x) e_i(x)\mathds{1}\{i \in \text{top}k(p(x),k)\}, %i \in \text{top}k\}%
\vspace{-3pt}
\end{equation}
where $p(\cdot)$ is the posterior vector of the gating network and $\text{top}k(\cdot)$ selects indices of the $k$ highest values.
It has reduced significantly computational complexity for the inference, compared to the MoE-based transformer structure; computational cost remains the same regardless of the number of experts.
They also have introduced an auxiliary training loss encouraging balanced usage of every expert, to efficiently leverage the network capacity and prevent biased training on a certain expert.
The detailed architecture is shown in~\Fig{modules} (a), where there is an additional general expert for the skip connection.
In this paper, we reform the sparse MoE-based transformer suitable for the multi-lingual ASR.
We will simply refer to the aforementioned network as MoE in this paper.

\begin{figure*}[t]
\centering
     \begin{subfigure}[b]{0.28\linewidth}
         \centering
         \includegraphics[width=\textwidth]{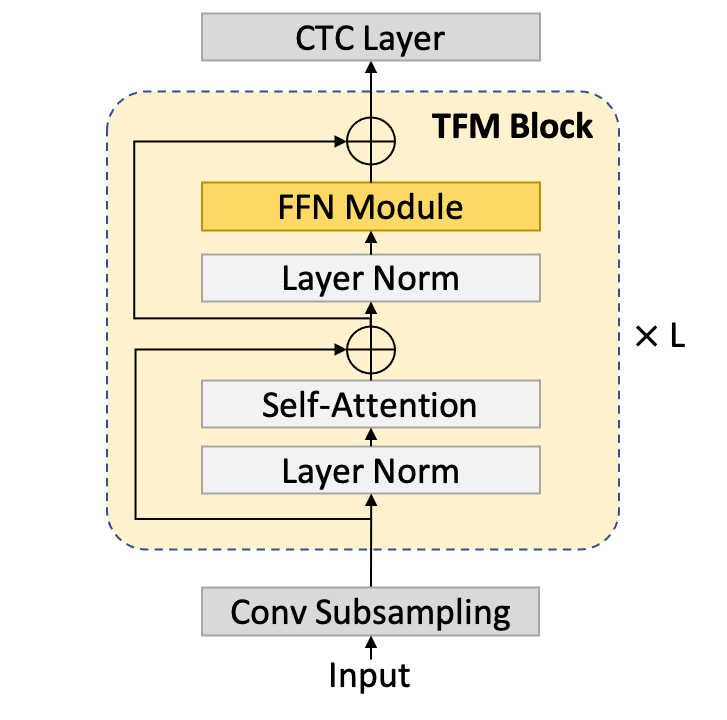}
         \caption{Vanilla Transformer Model}
\vspace{-5pt}
         \label{fig:TFM}
     \end{subfigure} 
     \begin{subfigure}[b]{0.28\linewidth}
         \centering
         \includegraphics[width=\textwidth]{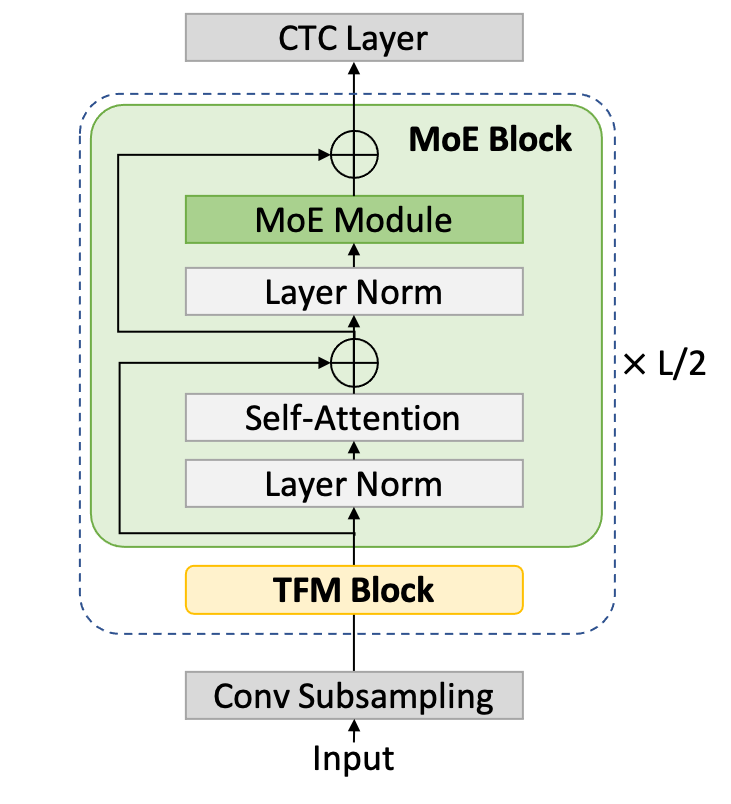}
         \caption{MoE-based Transformer Model}
\vspace{-5pt}
         \label{fig:MoE}
     \end{subfigure}    
     \begin{subfigure}[b]{0.28\linewidth}
         \centering
         \includegraphics[width=\textwidth]{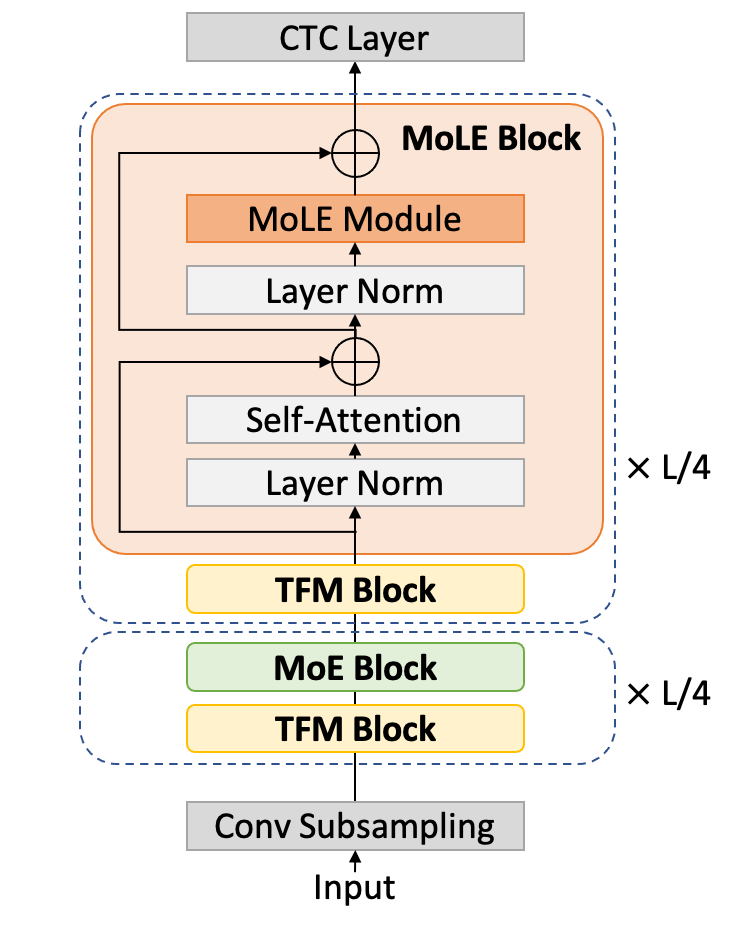}
         \caption{MoLE-based Transformer Model}
\vspace{-5pt}
         \label{fig:MoLE}
     \end{subfigure} 
     \caption{Overview of Transformer Blocks for CTC-ASR}
     \label{fig:overview}
    \vspace{-7pt}
\end{figure*}
\section{Proposed Method}
\label{sec:proposed}
\vspace{1pt}
We demonstrate our proposed model, Mixture-of-Language-Experts~(MoLE), for multi-lingual ASR, with several modifications on the MoE structure to better suit the task.
Instead of homogeneous experts in the MoE, the MoLE comprises two types of experts with their different roles: a language-specific expert and a language-agnostic expert.
% \jee{Different to MoE where homogeneous experts model both linguistic and acoustic information, the proposed MoLE comprises two types of experts with two different roles; a language-specific expert and a language-agnostic expert.}
The overall architecture of MoLE is presented in \Fig{modules} (b).

\vspace{-3pt}
\subsection{Language-specific Expert~(LsE)}
\vspace{-3pt}

We design the language-specific expert which perceives speech signals spoken in a specific language.
Thus, we employ $N$ LsEs in a MoLE layer to perceive $N$ languages.
The computational complexity of the MoLE does not increase regardless of the number of experts, due to the nature of the sparse MoE structure.
The major difference compared to the sparse MoE lies in the gating network.
First, our gating network is an LSTM-based model, ingesting the entire input sequence with sequential modeling for a \textit{global} token.
The utterance-wise global token taken from the last frame assigns the input sequence onto a specific expert, while the MoE routes input sequences using framewise tokens, which loses consistency.
Since an utterance is usually spoken in one language, it is not appropriate to activate different experts for every input speech frame.
Second, instead of the balancing loss in the MoE, we train the gating network using a language representation loss~(LRL) inspired in~\cite{snell2017prototypical}.
The language representation loss builds $N$ clusters of speech embeddings and minimizes distances between the centroid of the cluster and belonging embeddings while maximizing distances with the centroids of other languages.
The language representation loss $\mathcal{L}_{LR}$ is as follows,
\vspace{-3pt}
\begin{equation}
    \begin{split}
    \label{eq:proto_loss}
        \mathcal{L}_{LR}=-\text{log}p(k=n|z_r)=-\text{log}{e^{cos(z_r, c_n)} \over \sum_{n'} e^{cos(z_r, c_{n'})}}, \\
        \text{where}\quad c_n = {1 \over |S_n|}\sum_{z_r \in S_n}z_r
    \end{split}
\vspace{-3pt}
\end{equation}
where $z_r$ indicates hidden embedding of a gating network, $c_n$ is the centroid of hidden embedding which belongs to $n$-th language, and $S_n$ means a subset of speech spoken by $n$-th language.
LRL provides guidance on input signals to be assigned to the same expert if they are in the same language.
We introduce a metric-based learning strategy to benefit from stable training even in an imbalanced training set, rather than using discriminative training criteria (\eg cross-entropy).
Thus, the global tokens are chosen according to the language information in speech signals, and the gating network activates the same expert consistently for speech signals spoken in the same language.

\begin{table}[t]
\centering
\caption{Details of train and test datasets}
\vspace{-7pt}
\begin{tabular}{l |C{0.6cm} C{0.6cm} C{0.6cm} C{0.6cm} C{0.6cm} | R{0.6cm} } 
    \toprule
    \textbf{Dataset} & \textbf{En} & \textbf{Es} & \textbf{Fr} & \textbf{Zh} & \textbf{Ja} & \textbf{Total} \\
    \midrule
    Train (h) & 600 & 320 & 400 & 122 & 43 & 1,485 \\
    Test (h)  &  26 &  26 &  26 &  17 &  6 & 101 \\
    \bottomrule
\end{tabular}
\label{tab:data}
         \vspace{-10pt}
\end{table}

\vspace{-3pt}
\subsection{Language-agnostic Expert~(LaE)}
\vspace{-3pt}
In addition to LsEs, the MoLE consists of another expert, named language-agnostic expert~(LaE).
This expert ingests input data, agnostic to language types in the input speech.
The main role of LaE is to analyze acoustic characteristics not related to languages such as speech power, pitch, speaking environments, or speaker identity.
The normalization of these features for ASR is necessary, and it should be activated equally regardless of language.
Therefore, it is efficient to share an expert for all languages.
The other role of LaE is supporting LsEs by re-calibrating the outputs of MoLE.
We designed this module to supplement LsEs in case of low confidence $\gamma$ in gating decisions.
The highest posterior value $\gamma$ of gating networks is weighted to the outputs of the selected LsE, but the value is significantly variable about input signals (primarily influenced by the confidence of the gating network).
% Consequently, the variable dynamic range of expert outputs causes unstable learning of the model and its inference.
Here, we intend to compensate for the low confidence decision of the gating network by relatively increasing the weights of LaE outputs.
% Here, we adjust this problem by calibrating the output of LaE.
Thus, we calibrate the output of LaE by $(1-\gamma)$, and it stabilizes the outputs of MoLE. %which stabilizes the dynamic range of MoLE.
It is particularly advantageous on the multi-lingual ASR since the LaE intervenes more in the acoustic modeling, in case of low confidence on the gating network output.
\section{Experiments}
\label{sec:exp}
\subsection{Implementation Details}
\vspace{-3pt}

\noindent\textbf{Datasets} The dataset used in experiments is a subset of CommonVoice which contains 5 languages: English~(\textbf{En}), Spanish~(\textbf{Es}), French~(\textbf{Fr}), Chinese~(\textbf{Zh}) and Japanese~(\textbf{Ja}).
The total length of training data and test data are 1,485 hours and 101 hours, respectively, where the amount of each language set is not in balanced.
The amount of training and evaluation data of each language is depicted in~\Tab{data}.
In a training mini-batch, we randomly placed speech samples from the entire training set to follow the language distributions of the dataset.
The speech signal is transformed to 80-dimensional log-Mel filter-bank energy extracted at every 10ms with 25ms frame length in 16,000Hz sampling rate.

\begin{table*}[t]
\label{table:results}
\centering
\caption{Experimental results for CTC-ASR. CER of baseline, and the proposed method. $\mathcal{L}_{LR}$: language representation loss, \textbf{LaE}: language agnostic expert, \textbf{Calib.}: calibration of LaE.}
\vspace{-7pt}
\begin{tabular}{C{1cm} | C{1cm} || c |c c || C{1.1cm} | C{1.1cm} | C{1.1cm} | C{1.1cm} | C{1.1cm} | C{1.1cm} } 
    \toprule
    \textbf{Model}  & \textbf{Lang.} & $\mathcal{L}_{LR}$ & \textbf{LaE} & \textbf{Calib.} & \textbf{En} & \textbf{Es} & \textbf{Fr} & \textbf{Zh} & \textbf{Ja} & \textbf{OVR} \\
    \midrule
    TFM  & Mono &&&  & 15.20 & 5.82 & 8.86 & 17.82 & 9.54 & 10.32 \\
    TFM  & Multi &&&  & 18.08 & 8.33 & 12.24 & 19.88 & 17.65 & 13.34 \\
    \midrule
    MoE & Multi & & & & 17.50 & 7.41 & 11.20 & 18.59 & 15.64 & 12.43 \\
        &  & & \checkmark& & 15.84 & 6.48 & 9.81 & 17.60 & 13.56 & 11.10 \\
    \midrule
    MoLE & Multi &\checkmark& & & 17.06 & 6.94 & 10.43 & 17.20 &11.07 & 11.68 \\

    & & & \checkmark&& 15.79 & 6.20 & 9.62 & 16.84 & 10.12 & 10.82 \\
        & &\checkmark&\checkmark&         &15.51 & 6.30 & \textbf{9.27} & \textbf{15.99} & 9.96 & 10.59 \\
    
        & &\checkmark&\checkmark &\checkmark& \textbf{15.50} & \textbf{6.10} & 9.35 & 16.18 & \textbf{9.26} & \textbf{10.51} \\
    \bottomrule
\end{tabular}
\label{tab:results}
\vspace{-7pt}
\end{table*}

\noindent\textbf{Architecture}
We implemented 3 CTC-ASR models containing 12 transformer~(TFM) blocks, where the block settings are customized according to the model.
The baseline model follows similar architecture of~\cite{lahiri2021multilingual} with some modifications for a fair comparison.
It contains TFM block in every layer, containing a self-attention module and a feed-forward network~(FFN) in a TFM block (\Fig{overview} (a)).
% The basic model contains 12 transformer layers in~\Fig{overview} (a), and the other model contains MoE blocks.
The MoE-based transformer model placed MoE blocks after every TFM block as in~\Fig{overview} (b). %that consists of FFN experts and tokenizer
% The proposed model, the MoLE-based transformer model, has 3 stacks of a TFM and a MoE block on the bottom, 3 stacks of a TFM and a MoLE block on the top of the model, as shown in~\Fig{overview} (c).
The proposed model has MoLE modules replacing MoE modules, on the 8, 10, and 12th layers, as shown in \Fig{overview} (c).
Both in MoE and MoLE modules, each expert including LsE and LaE consists of 2 layers of FFN, which is the same as the FFN in the TFM block.
For a fair comparison with our model, we placed a common expert similar to LaE, where it is not regulated by the gating networks, in the MoE-based model.
% \yh{Every expert in MoE and MoLE consists of 2 layers of FFN including LsE, LaE.}
The type of gating network in MoLE is LSTM, whereas the MoE block has 2 fully-connected layers as the gating network.
% Every expert in our experiments shares the structural configuration as same as the FFN module in the TFM block.
% All experts in our experiments have the same architecture as the basic FFN module in TFM block.
Each self-attention layer has 256 dimension with 4 heads, and
FFN module has a hidden layer and a projection layer with 2,048 and 256 dimension, respectively.
The gating network consists of an LSTM with 64 cells and an fully-connected layer of 5 dimension same as the number of languages. 
The S2S decoder has 4 TFM layers of 2,048 dim.
For the ASR, we prepared the 7,000 word-piece model in total following~\cite{kudo2018sentencepiece}.
% The base MoE model is 6 blocks of MoE block like ~\Fig{overview} (b), and MoLE replaces the top 3 MoE blocks with MoLE blocks.
% The tokenizer in MoLE is made up of LSTM and FFN, while basic MoE is stacked FFN layers.
% And LRL is calculated by the mean of each layer's loss and 0.1 weight is used for proposed experiments.

\begin{table}[t]
\centering
\caption{Experimental results for S2S-ASR}
\vspace{-7pt}
\begin{tabular}{l || c | c | c | c | c | c} 
    \toprule
    \textbf{Model}  & \textbf{En} & \textbf{Es} & \textbf{Fr} & \textbf{Zh} & \textbf{Ja} & \textbf{OVR} \\
    \midrule
    TFM & 14.89 & 5.46 & 8.63 & 18.90 & 15.48 & 10.37 \\
    \midrule
    MoE   & 13.61 & 4.84 & 7.76 & 17.57 & 11.01 & 9.29 \\
    \midrule
    MoLE         & 13.74 & 5.03 & 7.74 & 17.82 & 10.35 & 9.37 \\
    \hspace{5pt}+Calib.   & \textbf{13.58} & \textbf{4.71} & \textbf{7.56} & \textbf{17.01} & \textbf{8.56} & \textbf{9.05} \\
    \bottomrule
\end{tabular}
\label{tab:s2s}
\vspace{-8pt}
\end{table}

\vspace{-3pt}
\subsection{Experimental Results}
\vspace{-3pt}
We compare performances between 3 CTC-ASR models in the character-error-rate~(CER).
We also analyze the impact of each training strategy through an ablation study.

In~\Tab{results}, we report the overall ASR performances of baseline models and the proposed model.
Compared to the mono-lingual ASR where one model learns a single language, multi-lingual ASR results showed degraded performances in every language since the model confuses vocabulary units that have similar pronunciation across languages. 
% Compared to the mono-lingual ASR where one model learns ASR in a single language, multi-lingual ASR results showed degraded performances in most languages.
% Because it is confusing for the model to distinguish a correct token from other multi-lingual tokens which sound similar.
In contrast, MoE and MoLE-based models showed overall improvements compared to the vanilla TFM model.
Moreover, the proposed MoLE even surpassed the mono-lingual ASR in Zh and Ja which are low-resource languages, since it lessened the lack of data and confusion of vocabulary by general acoustic modeling and LsE.
%since it lessened the lack of data for general acoustic modeling.
%utilized the model capacity efficiently.

Between MoE and MoLE models without the LaE, we found that leveraging LsEs instead of usual experts was significantly effective.
The increment in performance was especially noticeable in the case of low-resource languages.

\noindent\textbf{Ablation study} 
We conducted several ablation studies to verify the effectiveness of language representation loss ($\mathcal{L}_{LR}$), language-agnostic expert~(LaE), and the calibration on the LaE.
% In both MoE and MoLE models, introducing LaE increased the overall performance.
In both MoE and MoLE models, introducing LaE reduced the CER with more than 1\% in most cases.
% Especially in the low-resource languages, the MoLE model showed dramatic improvements.
The LRL brought slight improvements in overall results by stabilizing the gating network for consistent expert activation.
Also, the calibration on the LaE outputs with the residual of gating network weights resulted in lower CER, especially in Ja.

\vspace{-3pt}
\subsection{Sequence-to-sequence~(S2S)-ASR}
\vspace{-3pt}
To generalize the experimental results for our proposed model, we conducted another experiment, the sequence-to-sequence recognition model.
Similar to CTC-ASR experiments, we implemented TFM-based, MoE-based, and MoLE-based transformer in the encoder, respectively and there are 4 layers of MoE-based transformer with a cross-attention in the decoder following to~\cite{dong2018speech}.
In \Tab{s2s}, we reported the recognition in CER metric, and the MoE and MoLE approaches showed improved results compared to the TFM.
Comparing the MoE and MoLE, we found similar results of CTC-ASR, where there was performance improvement on our proposed model.
Specifically, the calibration of LaE led to improvements in the low-resource languages.

\section{Conclusion}
\label{sec:conclusion}
% \vspace{-2pt}
This work proposed a powerful multi-lingual ASR model, called Mixture-of-Language-Experts~(MoLE).
We introduced several modifications to the mixture-of-experts approaches, to be favorable to the multi-lingual ASR task.
% The global token from
% The gating network assigns an input sequence to a certain language-specific expert according to the language type, and the language-agnostic expert perceives it regardless of language.
The gating network assigns input sequence to a certain language-specific expert by sentence according to the language type, and the language-agnostic expert perceives it regardless of language.
% Language representation loss is proposed for implicit language recognition and enhancing the ability of routing.
Experiments for CTC-ASR and S2S-ASR verified the superiority of our proposed method with a significant improvement in multi-lingual recognition, especially in low-resource languages. 
In addition, we conducted several ablation studies to examine our strategies for the MoLE, which led to meaningful recognition improvements.
% Experiments for CTC-ASR and S2S-ASR demonstrated that the proposed method achieved a significant improvement in multi-lingual settings, especially in low-resource language sets.
% In future work, we are going to work on extending the number of experts by referring to~\cite{xie2022moec} and exploring the method to apply other frameworks.

\vfill\pagebreak

% \section{REFERENCES}
% \label{sec:refs}

\bibliographystyle{IEEEbib}
\bibliography{shortstrings,refs}

\vfill\pagebreak
\end{document}